\documentclass[prl,twocolumn,superscriptaddress]{revtex4}
\usepackage{graphicx}
\usepackage{url}
\newcommand{\etal}{{\it et al.}}

\begin{document}

\title{Quantum oscillations in a biaxial pair density wave state}

\author{M. R. Norman}
\affiliation{Physical Sciences and Engineering Directorate, Argonne National Laboratory, Argonne, IL 60439}
\author{J. C. S\'{e}amus Davis}
\affiliation{LASSP, Department of Physics, Cornell University, Ithaca, NY 14853}
\affiliation{Condensed Matter Physics Department, Brookhaven National Laboratory, Upton, NY 11973}

\begin{abstract}
There has been growing speculation that a pair density wave state is a key component of the phenomenology of
the pseudogap phase in the cuprates.  Recently, direct evidence for such a state has emerged from an analysis
of scanning tunneling microscopy data in halos around the vortex cores.  By extrapolation, these vortex halos would then 
overlap at a magnetic field scale where quantum oscillations have been observed.  Here, we show that a biaxial pair density wave
state gives a unique description of the quantum oscillation data, bolstering the case that the pseudogap
phase in the cuprates may be a pair density wave state.
\end{abstract}

\date{\today}

\maketitle

The discovery of charge density wave correlations in cuprates by neutron and x-ray scattering, scanning tunneling
microscopy (STM), and nuclear magnetic resonance has had a
profound influence on the field of high temperature superconductivity, but a number of observations indicate that the
cuprate pseudogap phase involves more than just
charge ordering \cite{rmp1}.  Evidence for pairing correlations, as well as time reversal symmetry breaking, is
apparent depending on the particular experimental probe.  In an attempt to make sense of various conflicting interpretations, it was
speculated that a pair density wave (PDW) state,
evident in numerical studies of the $t-J$ and Hubbard models \cite{corboz,zheng},
could be the `mother' phase and also gives
a natural explanation for angle resolved photoemission data \cite{lee}.
More direct evidence has emerged from STM using a superconducting tip, where it was shown that the pairing order parameter was
modulated in space \cite{hamidian}.  But more telling evidence has recently come from looking at scanning tunneling
data in a magnetic field \cite{edkins}.  There, direct evidence was found for biaxial order in a halo surrounding the vortex
cores at a wave vector that was one-half that of the charge density wave correlations, exactly as expected based on PDW 
phenomenology \cite{pdw2}.
This last observation leads to an obvious conjecture.  One can estimate the field at which these vortex halos overlap \cite{wu,foot1},
and this field is the same at which a long range ordered charge density wave state has been seen by x-rays scattering \cite{gerber}.
Interestingly, this is virtually the same field at which quantum oscillations also become evident \cite{dhva,dhva2}.  This implies that the small electron pockets
inferred from these data are due to the state contained in these vortex halos.

The most successful model for describing quantum oscillation data is that of Harrison and Sebastian \cite{neal}.  By assuming
a biaxial charge density wave state, they are able to form nodal pockets by folding of the Fermi arcs observed by photoemission
to form an electron diamond-shaped pocket centered on the $\Gamma$-point side of the Fermi arc observed by angle resolved
photoemission \cite{arpes}.  In their scenario, as this pocket grows, eventually a Lifshitz
transition occurs, leading to a hole pocket centered around the $\Gamma$ point itself.  A central question is whether an alternate
model could have a similar phenomenology.

To explore this issue, we consider a biaxial PDW state \cite{foot2} with a wave vector of magnitude $Q=\pi/4a$ as observed in the recent 
STM data \cite{edkins}.  The secular matrix for such a state is of the form:
\begin{eqnarray}
\left(\begin{array}{ccccc}
\epsilon_{\vec{k}} & \Delta_{\vec{k}+(\frac{Q}{2},0)} & \Delta_{\vec{k}-(\frac{Q}{2},0)} & \Delta_{\vec{k}+(0,\frac{Q}{2})} & \Delta_{\vec{k}-(0,\frac{Q}{2})} \\ 
\Delta_{\vec{k}+(\frac{Q}{2},0)} & -\epsilon_{-\vec{k}-(Q,0)} & 0 & 0 & 0 \\ 
\Delta_{\vec{k}-(\frac{Q}{2},0)} & 0 & -\epsilon_{-\vec{k}+(Q,0)} & 0 & 0 \\ 
\Delta_{\vec{k}+(0,\frac{Q}{2})} & 0 & 0 & -\epsilon_{-\vec{k}-(0,Q)} & 0 \\
\Delta_{\vec{k}-(0,\frac{Q}{2})} & 0 & 0 & 0 & -\epsilon_{-\vec{k}+(0,Q)} \nonumber
\end{array}\right)
\label{secular}
\end{eqnarray}
Here, we assume a d-wave form for the PDW order parameter, $\Delta_{\vec{q}} = \frac{\Delta_0}{2}(\cos(q_xa)-\cos(q_ya))$, with its argument,
$\vec{q}=\vec{k}+\frac{\vec{Q}}{2}$, being the Fourier transform of the
relative coordinate of the pair (the center of mass Fourier transform being $\vec{Q}$).  We also ignore all of the other off diagonal components,
which arise from the secondary charge order, as they only lead to quantitative corrections to the results presented here.
For $\epsilon_{\vec{k}}$ we assumed the tight binding dispersion of He \etal~\cite{he} for Bi2201.  We do this for two reasons.  First, this was the
dispersion considered in previous work on PDWs \cite{lee}.  Second, there are no complications in this dispersion associated with bilayer splitting.

To proceed, we need to define the spectral function, $A$, as measured by angle resolved photoemission:
\begin{equation}
A(\omega,\vec{k}) = \frac{\Gamma}{\pi}\frac{c_i(\vec{k})^2}{(\omega-E_i(\vec{k}))^2+\Gamma^2}
\end{equation}
Here, $E_i$ is the i'th eigenvalue of the secular matrix, $c_i$ the $\vec{k}$ component of the corresponding eigenvector
(the analogue of the particle-like Bogoliubov component), and $\Gamma$
a phenomenological broadening parameter.

\begin{figure}
\includegraphics[width=\hsize]{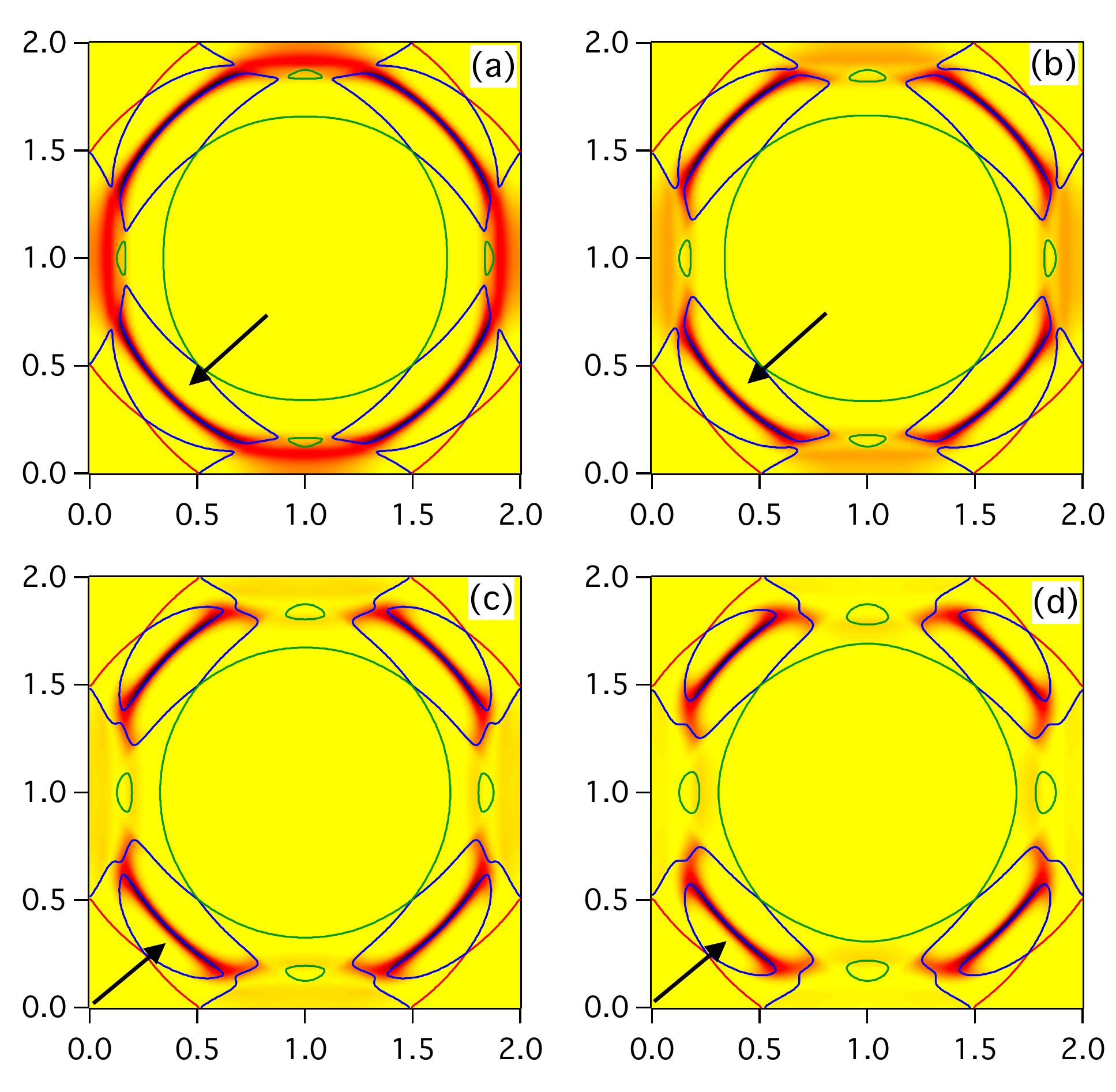}
\caption{Spectral weight and eigenvalue contours at $\omega=0$ for a pair density wave state with its amplitude, $\Delta_0$,
being (a) 25 meV, (b) 50 meV, (c) 75 meV and (d) 100 meV (the x and y axes are $k_x$ and $k_y$ in units of $\pi/a$).
Arrows point to the center of the electron pocket ((c) and (d)) and the hole pocket ((a) and (b)).
The normal state dispersion is given by He \etal~\cite{he}.  
Here, the modulus of the PDW ordering vector, $Q$, is $\pi/4a$, as observed in recent STM experiments \cite{edkins}.  For
the spectral weight, a phenomenological broadening parameter, $\Gamma$, of 25 meV is assumed.}
\label{fig1}
\end{figure}

In Fig.~1, we show the spectral weight and eigenvalue contours at $\omega=0$ for four values of $\Delta_0$.   Deep in the pseudogap
phase (large $\Delta_0$), a small electron pocket centered along the diagonal $(0,0)-(\pi,\pi)$ is observed whose flat edge follows the 
spectral weight. 
As such, this pocket should dominate
the deHaas-vanAlphen (dHvA) amplitude, unlike the other pockets which exhibit no spectral weight \cite{foot3}.
 As the hole doping increases (smaller $\Delta_0$), this pocket undergoes a Lifshitz transition,
resulting in a larger hole pocket also centered along the diagonal that resembles that obtained in the phenomenological YRZ
(Yang-Rice-Zhang) model for the cuprates \cite{yrz,yang}.  Once the gap collapses, then one recovers the much larger hole
pocket centered at $(\pi/a,\pi/a)$ that is characteristic of the overdoped state \cite{Tl2201}.
We remark that the biaxial order is critical in forming these smaller pockets, though hints of them can be found in earlier work that assumed a
uniaxial PDW instead \cite{berg,loder,zelli,zelli2} (the last two of these papers addressing the dHvA data).

\begin{figure}
\includegraphics[width=\hsize]{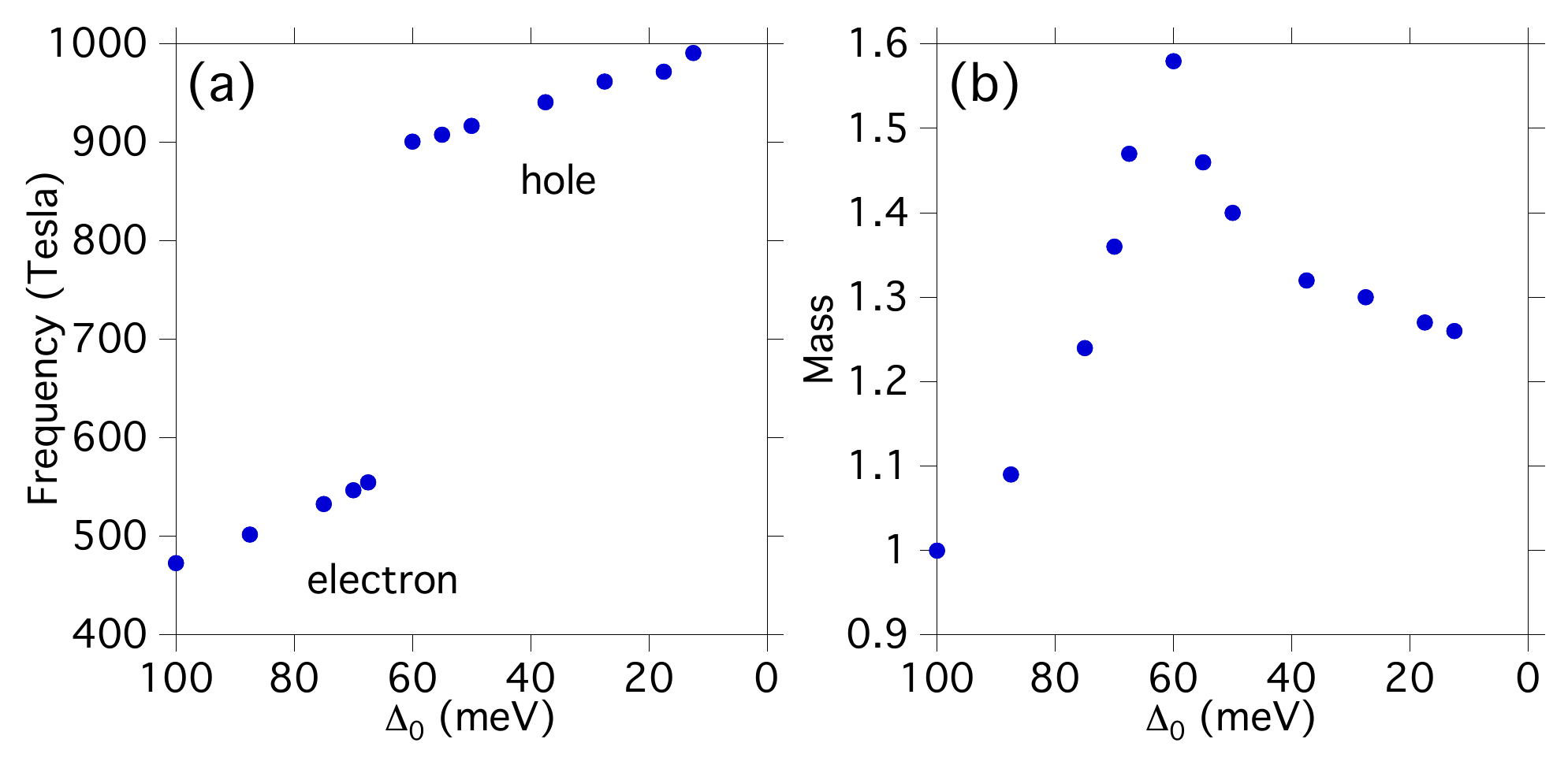}
\caption{(a) deHaas-vanAlphen (dHvA) frequency (area of the pocket) and (b) cyclotron mass versus $\Delta_0$ as derived from Fig.~1.
Note the Lifshitz transition where the small electron pocket for large $\Delta_0$ converts to a larger hole pocket for smaller $\Delta_0$.
As a reference, the area of the large normal state hole pocket centered at $(\pi/a,\pi/a)$ for $\Delta_0=0$ is 17.68 kilo-Tesla with a cyclotron mass of 3.69.}
\label{fig2}
\end{figure}

We quantify this by plotting the area of the pocket (in the dHvA units of Tesla) along with the cyclotron mass as a function of $\Delta_0$ in
Fig.~2.  One sees a modest dependence of the pocket area on $\Delta_0$ except for the pronounced jump at the Lifshitz transition,
along with the associated mass divergence at the Lifshitz transition.  These dependencies are in good accord with dHvA data as a
function of hole doping \cite{mass}, including the mass divergence, noting that quantitative details are influenced by the dispersion and 
chemical potential (that is, the conversion of the x-axis of Fig.~2 to doping is influenced not only by the doping dependence of $\Delta_0$, but
also by the doping dependence of the band structure and chemical potential).  Moreover, the results presented here offer a prediction.  That is, 
beyond the mass divergence (as $\Delta_0$ decreases),
there should be a small doping range where a large hole pocket of roughly twice the size of the electron pocket occurs before the very large
hole pocket in the overdoped regime forms when the gap collapses.
This prediction is supported by Hall data that shows a region of the phase diagram between $p$=0.16 and $p$=0.19 where the
Hall constant rapidly changes \cite{badoux}, with $p$=0.16 being where the mass divergence referred to above occurs, and $p$=0.19 where the large Fermi
surface is recovered (here, $p$ is the doping).

We feel that the scenario offered here is an attractive alternate to models based on a charge density wave.  It is not only consistent with recent STM
data in the vortex halos \cite{edkins}, but also consistent with magneto-transport data that indicate the presence
of pairing correlations for magnetic fields not only up to but well beyond the resistive H$_{c2}$ \cite{hsu}.
This is in line as well with previous theoretical work on quantum oscillations in a d-wave vortex
liquid \cite{banerjee}.  Certainly, we hope that the model offered here will lead to additional studies in high magnetic fields to
definitively determine whether a pair density wave state really exists.

In summary, the work presented here bolsters the case that the enigmatic pseudogap phase in the cuprates is a pair density wave state.

This work was supported by the Center for Emergent Superconductivity, an Energy Frontier Research Center funded by the
US DOE, Office of Science, under Award No.~DE-AC0298CH1088.
We thank Stephen Edkins, Mohammad Hamidian and Andrew Mackenzie for access to their vortex halo STM data in advance of publication.
We also acknowledge Neil Harrison, Peter Johnson, Marc-Henri Julien, Catherine Kallin, Steve Kivelson, Patrick Lee, Brad Ramshaw, Subir Sachdev, 
Suchitra Sebastian, Todadri Senthil, and Louis Taillefer for various discussions.


\begin{thebibliography}{99}
\bibitem{foot1}
The field at which the area of the halos is equal to the total area is $\Phi_0/(\pi\xi^2)$ where $\Phi_0$ is the flux quantum and $\xi$ the radius
of the halo.  For a typically observed radius of 50~$\AA$, this field would be 26 Tesla.
\bibitem{foot2}
Here, we assume a homogeneous PDW order parameter, and ignore any uniform superconducting
component, along with any effects due to vortices and their associated phase winding.  This should 
be justified for the high fields at which the quantum oscillations are observed.
\end{thebibliography}
\end{document}